\documentstyle[12pt]{article}
\begin{document}
\begin{flushright}
DTP--MSU 95/39\\
October 1995
\end{flushright}
\vskip4cm\begin{center}
{\LARGE\bf
Geroch--Kinnersley--Chitre group for Dilaton--Axion Gravity}\footnote
{A talk given at the International Workshop, 
``Quantum Field Theory under the Influence of External Conditions'', Leipzig
Germany, 18--22 September 1995, published in the Proceedings,  
M. Bordag (Ed.), B.G. Teubner Verlagsgessellschaft, Stuttgart--Leipzig, 
1996, pp. 228-237.
}\\ \vskip1cm
{\bf
D.V. Gal'tsov} \\
\normalsize  Department of Theoretical Physics,  Moscow State University,\\ 
\normalsize Moscow 119899, {\bf Russia}\\
  \vskip2cm
{\bf Abstract}\end{center}
\begin{quote}
Kinnersley--type representation is constructed for the four--dimensional 
Einstein--Maxwell--dilaton--axion system restricted
to space--times possessing two non--null commuting Killing symmetries.
New representation essentially uses the matrix--valued $SL(2,R)$ formulation
and effectively reduces the construction of the Geroch
group to the corresponding problem for the vacuum Einstein equations.
An infinite hierarchy of potentials is introduced in terms of $2\times 2$
real symmetric matrices directly generalizing the scalar
hierarchy of Kinnersley--Chitre known for the vacuum Einstein equations. 
 \vskip5mm
\noindent
PASC number(s): 97.60.Lf, 04.60.+n, 11.17.+y \vskip1cm
\end{quote}
\newpage

``Geroch--Kinnersley--Chitre group'' is a popular name for 
the infinite symmetry arising
in the Einstein and Einstein--matter theories enjoing a complete
integrability property after reduction to two dimensions via imposition
of a rank two Abelian isometry group on the four--dimensional space-time.
An original observation by Geroch \cite{ge}, essentially based on the                              
earlier works of Ehlers \cite{eh} and Harrison \cite{ha}, was then
developed by Kinnersley \cite{ki1} and Kinnersley
and Chitre \cite{kichi}. Important insights in establishing the
integrability of the two--dimensional Einstein and Einstein--Maxwell (EM)
theories were due to the works of Neugebauer and Kramer \cite{ne}, 
\cite{nk} in which the relevance of the three--dimensional $\sigma$--models
on symmetric spaces has been discovered. A concise complex potential
representation given by Ernst \cite{er} for both Einstein and 
Einstein--Maxwell theories provided new structures particularly useful 
in this research. Different proofs of the complete integrability
of two--dimensional reductions of the Einstein and EM systems were given
in the late 70--ths by Harrison \cite{har}, Maison \cite{ms}, 
Belinskii and Zakharov \cite{bz}, and Hauser and Ersnt \cite{her}. These
alternative formulations as well as some other approaches 
\cite{nb}, \cite{al1} were compared and extended by Cosgrove \cite{cg}.
Extensive application of related methods in General Relativity was
very fruitful in the search of exact classical solutions \cite{ex}
(for more recent development and review see \cite{gur}, \cite{ex2},
\cite{kn1}, \cite{bm}, \cite{le}, \cite{al2}, \cite{sb}, 
\cite{do}, \cite{vd}, \cite{gri}).

In more general and modern language, Geroch group may be seen
as an infinite--dimensional symmetry acting on harmonic maps
of Riemann surfaces into symmetric spaces, the associated methods
being related to the study of deformation of harmonic maps. In the
gravity context such maps are connected to the three--dimensional 
$\sigma$--models to which the Einstein and some Einstein--matter
systems are reduced for space--times admitting a non--null Killing
vector field. If the corresponding target space is a symmetric
riemannian space, its finite isometries undergo infinite affine
extension when one goes to two dimensions. Some other physically 
interesting systems possess
such a property, the most notable examples being the Kaluza--Klein
\cite{ne}, \cite{ma}, \cite{cl} and certain supergravity models 
\cite{ju}, \cite{bgm}, \cite{ni}. 

It is worth noting that similar
infinite symmetries are encountered in a  different class of theories,
two--dimensional conformal field models, in particular, in the 
theory of superstrings. In the low energy limit such theories lead
to the gravity--matter systems which may have in their turn the
associated infinite symmetries. The investigation of links between 
two theories seems to be a promising field of research. Here we discuss
Geroch--type symmetries found recently in the heterotic string 
low energy effective theory \cite{gk}, \cite{g}, \cite{bmss}. 
In the bosonic sector of this theory one finds
the Einstein gravity coupled to massless vector and scalar fields.
The simplest model of this kind incorporating basic features of the full 
effective action --- ``dilaton--axion gravity'', or 
Einstein--Maxwell--dilaton--axion (EMDA) system --- can be 
formulated directly in four dimensions where it includes one $U(1)$ vector 
and two scalar fields coupled in  such a way that the theory possesses
non--abelian $SL(2, R)$ duality symmetry. 
 
To begin, let us review briefly how Geroch symmetry emerges in
the vacuum Einstein theory. Consider a four--dimensional manifold 
admitting a two--parameter group of motions generated by the 
Abelian algebra of Killing vectors $K_A,\;\; A=1, 2$. In the adapted 
coordinates $K_A=\partial_A$ the interval can be written as
\begin{equation}
ds^2=f_{AB}dx^A dx^B-h_{MN}dx^{M+2}dx^{N+2},
\end{equation}
where $A, B, M, N=1, 2$ and $f_{AB},\;\; h_{MN}$ are real symmetric
$2\times 2$ matrices depending only on $x^3,\; x^4$. In the case of
stationary axisymmetric metrics which we choose here for definitness
a particular gauge (Lewis--Papapetrou) is appropriate:
\begin{equation}
f_{11}=f,\quad  f_{12}=-f\omega,\quad f_{22}=f\omega^2-\rho^2/f,
\end{equation}
and 
\begin{equation}
h_{11}=h_{22}=e^{2\gamma}f^{-1},
\end{equation}
where $f,\; \omega,\; \gamma$ are functions of $x^3=\rho,\;x^4=z$.
The resulting two--dimensional equations can be conveniently 
written down using the Hodge--conjugated operators
\begin{equation}
\nabla=\partial_{N+2}=(\partial_z,\;\;-\partial_rho),\quad
{\tilde \nabla}={\tilde\partial}_{N+2}=(-\partial_\rho,\;\;\partial_z).
\end{equation}

The vacuum Einstein equations lead to the following system of
equations for $f_{AB}$
\begin{equation}
\nabla \left(\rho^{-1}f^{AB}\nabla f_{BC}\right)=0,
\end{equation}
where rising and lowering of indices is effected using the alternating
symbol
\begin{equation}
\epsilon_{AB}=-\epsilon^{AB}=
\left(\begin{array}{crc}
0&-1\\
1&0\\
\end{array}\right).
\end{equation}
Once $f_{AB}$ is found, the function $\gamma$ can be constructed by solving
a simple system of partial differential equations of the first order
(what remains true for the EMDA system too, so we will not concentrate
on this problem here).

These equations can be shown to possess two finite symmetry groups. 
The first is the so--called Matzner-Misner $SL(2, R)$ group $G$
\begin{equation}
x^A\rightarrow {G^A}_B\;x^B,\quad \det G=1,
\end{equation}
under which $f_{AB}$ transforms as the $SL(2, R)$ tensor
\begin{equation}
f_{AB}\rightarrow {G_A}^C{G_B}^Df_{CD}.
\end{equation}
The second, known as the Ehlers group $H$ (also $SL(2, R)$), acts 
on the initial metric variables in a non--local way. To make this explicit
one introduces a twist potential $\chi$, the existence of which is implied 
by the $11$--component of the Eq. (5), via duality relation
\begin{equation}
{\tilde\nabla}\chi=\rho^{-1} f^2\nabla \omega. 
\end{equation}
The set of variables $f,\;\chi$ gives rise to another real symmetric 
matrix $m_{AB}$,
\begin{equation}
m_{11}=f^{-1},\quad m_{12}=\chi f^{-1},\quad m_{22}=\chi^2 f^{-1}+f,
\end{equation}
in terms of which the equivalent to (5) set of equations reads
\begin{equation}
\nabla \left(\rho \;m^{AB} \;\nabla m_{BC}\right)=0.
\end{equation}
This tensor transforms nonlinearly under the Ehlers group consisting of
a gauge ($\chi\rightarrow \chi+\lambda_g$), scale
($f,\chi\rightarrow e^{2\lambda_s}f,\;e^{2\lambda_s}\chi$) and
a proper Ehlers transformation, which can conveniently be expressed
in terms of the Ernst potential $E=if-\chi$ as
\begin{equation}
E\rightarrow \frac{E}{1+\lambda_E E}.
\end{equation}
 
Acting by these transformations on some solution to the Eqs. (5) one 
obtains new solutions. Now, if $G$--covariance is taken into account,
one can construct new symmetries by conjugation of $H$ with $G:\;GHG^{-1}$.
Repeating this operation one is led to an infinite--dimensional group
\begin{equation}
K=...H\times G\times H\times G\times H,
\end{equation}
which is the Geroch--Kinnersley--Chitre group. It can be realised on an infinite
hierarchy of potentials  which may be introduced via the $G$--covariant
dualization procedure \cite{ki1}. 

The same reasoning was shown to hold for the EM system
(Kinnersley and Chitre \cite{kichi}) where the hierarchy is more
complicated and involves some additional structures. Remarkably, 
the EMDA system turns out to be in this respect simpler than the EM one,
and involves only the strustures typical for the vacuum Einstein
equations. 

Consider the EMDA action containig a metric $g_{\mu\nu}$, a $U(1)$
vector field $A_\mu$, a Kalb-Ramond antisymmetric tensor field
$B_{\mu\nu}$, and a dilaton $\phi$ in four dimensions
\begin{equation}
S=\frac{1}{16\pi}\int \left\{-R+\frac{1}{3}e^{-4\phi}
H_{\mu\nu\lambda}H^{\mu\nu\lambda}+
2\partial_\mu\phi\partial^\mu\phi -e^{-2\phi}F_{\mu\nu}
F^{\mu\nu}\right\}\sqrt{-g}d^4x,
\end{equation}
where
\begin{equation}
H_{\mu\nu\lambda}=\partial_{\mu} B_{\nu\lambda} -A_{\mu} F_{\nu\lambda}
\quad + \quad {\rm cyclic},
\end{equation}
and $F_{\mu\nu}=\nabla_\mu A_v-\nabla_\nu A_\mu$. After reduction to two
dimensions the set of dynamical quantities, in addition to metric variables
described above, contains two components of the 4--potential
\begin{equation}
A_B=\frac{1}{\sqrt{2}}(v,\;a),
\end{equation}
one non--trivial component of the Kalb--Ramond field
\begin{equation}
B_{12}=b,
\end{equation}
and the dilaton $\phi$. The corresponding system in presence of one
Killing vector field was studied in \cite{gk} and in the two--dimensional
case in \cite{g} in $G$--noncovariant way. Here we give $G$--covariant
formulation which opens a way to construct the Geroch--Kinnersley
group explicitly. From the results of \cite{gkt} it is clear that
the underlying $H$ group is a symplectic group $Sp(4, R)$, which fits 
well with the $SL(2,R)$ structure of the group G. One can find 
$G$--covariant description of the equations of motion simply by
introducing an additional $2\times 2$ matrix structure into the
Kinnesley formalism described above.

First we introduce the following two $2\times 2$ real symmetric matrices
instead of the one--component quantities $f$ and $\omega$ above:
\begin{equation}
{P}=
\left( \begin{array}{ccc}
f-e^{-2\phi}v^2 & -e^{-2\phi}v\\
-e^{-2\phi}v  & -e^{-2\phi}\\
\end{array} \right), \quad
{\Omega}=
\left( \begin{array}{ccc}
\omega & -q\\
-q  & qv-b\\
\end{array} \right)
\end{equation}
where $q=a+\omega v$.
Using them as building blocks, one can now construct a $G$--tensor
\begin{equation}
F_{AB}=
\left( \begin{array}{ccc}
P & -P\,\Omega\\
-\Omega \,P&\Omega \,P\, \Omega -\rho^2 P^{-1}\\
\end{array} \right).
\end{equation}
Raising and lowering the indices now has to be performed with
the matrix--valued alternating tensor
\begin{equation}
\epsilon_{AB}=-\epsilon^{AB}=
\left(\begin{array}{crc}
0&-I\\
I&0\\
\end{array}\right), \quad
I=
\left(\begin{array}{crc}
1&0\\
0&1\\
\end{array}\right).
\end{equation}
Note that the inverse tensor
\begin{equation}
{F^{-1}}_{AB}=\rho^{-2}F^{AB}=
\rho^{-2}\left( \begin{array}{ccc}
\rho^2\,P^{-1}-\Omega \,P \,\Omega & -\Omega\, P\\
-P\,\Omega&- P\\
\end{array} \right).
\end{equation}

In terms of this matrix the dynamical equations of motion assume
essentially the same form as (5):
\begin{equation}
\nabla \left(\rho^{-1}F^{AB}\nabla F_{BC}\right)=0.
\end{equation}
Four $2\times 2$ equations here are not algebraically independent.
As two independent equations the $11$ and $22$ ones 
can be conveniently chosen 
\begin{equation}
\nabla \left(\rho^{-1} P\nabla\Omega P\right)=0,
\end{equation}
\begin{equation}
\nabla \left(\rho\nabla P P^{-1} + \rho^{-1} P\nabla\Omega P\Omega\right)=0,
\end{equation}
The {\sc 22} component is related to (24) by transposition, while the $21$ one
is satisied automatically due to both (23) and (24).

Ehlers group for the EMDA system was first described in \cite{gk}
as some ten--parameter semisimple Lie group and then identified with
$Sp(4, R)$ in \cite{g}. It acts non--linearly on the set of six
variables consisting of three initial variables $f,\;v,\;\phi$
and three dualized variables $\chi,\;u,\;\kappa$. 
Here $\kappa$ is the pseudoscalar Pecci--Quinn counterpart to the
Kalb--Ramond field
\begin{equation}
H^{\mu\nu\lambda}=\frac{1}{2}e^{4\phi}E^{\mu\nu\lambda\tau}
\frac{\partial\kappa}{\partial x^\tau},
\end{equation}
$u$ is the magnetic potential related to the  Maxwell tensor
as
\begin{equation}
e^{-2\phi} F^{N2}+\kappa {\tilde F}^{N2}=
\frac{f^2 e^{-2\gamma}}{\sqrt{2}\rho}{\tilde \partial}_{N+2}u,
\end{equation}
while $\chi$ is the twist potential which now is the solution
of the equation
\begin{equation}
\nabla \chi=u\nabla v- v\nabla u-\rho^{-1} f^2{\tilde\nabla\omega}.
\end{equation}

From six real variables entering the problem 
as a set of scalar fields one can build 
three complex potentials of the Ernst type
\begin{equation}
\Phi=u-zv, \quad
E = if-\chi+v\Phi, \quad
z = \kappa +ie^{-2\phi}.
\end{equation}
It is worth noting that neither $\Phi$, nor $E$ reduce to the
original Ernst potential for Einstein--Maxwell system. However
their role in formulation of integrable system of equations is very
similar to that of the original Ernst potentials (more detailed discussion
can be found in \cite{gkt}). Now we have one more complex variable
(complex dilaton--axion field $z$) to describe the system. However
the theory has additional symmetries which effectively reduce
its complexity. One can show, in particular, that the whole theory
(not only with two, but also with one Killing symmetry imposed on
a four--dimensional manifold) is invariant under a discrete transformation
\begin{equation}
z^{'}=E,\quad  E^{'}=z,\quad \Phi^{'}=\Phi,  
\end{equation}
i.e. an interchange of $z$ and $E$. Within the EMDA
system the complex dilaton--axion field and the Ernst variable form
quite similar algebraic structures.

New dualized variables can be combined within the 
$2\times 2$ matrix $Q$ related to
$\Omega$ by matrix dualization \cite{gks}
\begin{equation}
\nabla Q=-\rho^{-1} \,P{\tilde \nabla} \,\Omega\,P.
\end{equation}
Explicitly it reads
\begin{equation}
Q=
\left( \begin{array}{ccc}
wv-\chi & w\\                  
w&-\kappa\\
\end{array} \right),
\end{equation}
where $w=u-\kappa v$. Using $P,\;Q$ pair one can now build $4\times 4$
matrix transforming as an $Sp(2, R)$ matrix--valued object which is 
a direct analog of the matrix $m_{AB}$ (10)
\begin{equation} 
M_{AB}=\left(\begin{array}{crc}
P^{-1}&P^{-1}Q\\
QP^{-1}&P+QP^{-1}Q\\
\end{array}\right).
\end{equation}
In new terms the equations of motion read
\begin{equation}
\nabla \left(\rho \;M^{AB} \;\nabla M_{BC}\right)=0.
\end{equation}

The $H$--transformations are most conveniently described in terms of the  
complex combination 
\begin{equation}
Z=Q+iP=
\left(\begin{array}{crc}
E&\Phi\\
\Phi&-z\\
\end{array}\right).
\end{equation}
They consist of the shift on the constant real symmetric matrix
\begin{equation}
Z \rightarrow Z+R,\qquad
R=
\left(\begin{array}{crc}
\lambda_g&\lambda_m\\
\lambda_m&-\lambda_{d_1} \\
\end{array}\right),\quad
\end{equation}
matrix ``scale'' transformation
\begin{equation}
Z \rightarrow S^TZS, \qquad
S=
\left(\begin{array}{crc}
e^{\lambda_s}&\lambda_{H_1}\\
-\lambda_e&e^{\lambda_{d_3}} \\
\end{array}\right),\quad
\end{equation}
and the shift of the inverted matrix 
\begin{equation}
Z^{-1} \rightarrow Z^{-1}+L,\qquad
L=
\left(\begin{array}{crc}
\lambda_E&\lambda_{H_2}\\
\lambda_{H_2}&-\lambda_{d_2} \\
\end{array}\right).
\end{equation}
The parameters $\lambda$ introduced here correspond to ten 
$H$--transformations (forming $Sp(4, R)$) generalizing the Ehlers
group to the EMDA system. It is worthwhile to list them separately.
Parameters $\lambda_g,\; \lambda_e,\; \lambda_m$ correspond to 
the gravitational, electric and magnetic gauge transformations:
\begin{equation}
E=E_0+\lambda_g,\quad \Phi=\Phi_0,\quad z=z_0,
\end{equation}
\begin{equation}
E=E_0-2\lambda_e \Phi_0-\lambda_e^2 z_0,\quad 
\Phi=\Phi_0+\lambda_e z_0,\quad z=z_0,
\end{equation}
\begin{equation}
E=E_0,\quad \Phi=\Phi_0+\lambda_m,\quad z=z_0.
\end{equation}
The scale transformation now reads
\begin{equation}
E=e^{2\lambda}E_0,\quad \Phi=e^{\lambda}\Phi_0,\quad z=z_0.
\end{equation}
The $SL(2, R)\;S$--duality subgroup is
\begin{eqnarray}
 E&=&E_0,\quad \Phi=\Phi_0,\quad z=z_0+\lambda_{d_1},\nonumber\\
E&=&E_0+\lambda_{d_2}\frac{\Phi_0^2}{1+\lambda_{d_2} z_0},\quad 
\Phi=\frac{\Phi_0}{1+\lambda_{d_2} z_0},\quad 
z=\frac{z_0}{1+\lambda_{d_2} z_0},\\ 
E&=&E_0,\quad \Phi=e^{\lambda_{d_3}}\Phi_0,\quad 
z=e^{2\lambda_{d_3}}z_0.\nonumber 
\end{eqnarray}
The most nontrivial part of the group in the physical terms
consists of the pair of the electric and magnetic Harrison 
transformations 
\begin{equation}
E=E_0,\quad \Phi=\Phi_0+\lambda_{H_1} E_0,\quad 
z=z_0-2\lambda_{H_1} \Phi_0-\lambda_{H_1}^2 E_0,
\end{equation}
\begin{eqnarray}
&&E=\frac{E_0}{(1+\lambda_{H_2}\Phi_0)^2+\lambda_{H_2}^2 E_0 z_0},\quad
\nonumber\\
&&\Phi=\frac{\Phi_0(1+\lambda_{H_2}\Phi_0)+\lambda_{H_2} E_0 z_0}
{(1+\lambda_{H_2}\Phi_0)^2+\lambda_{H_2}^2 E_0 z_0},\quad  \\
&&z=\frac{z_0}{(1+\lambda_{H_2}\Phi_0)^2+\lambda_{H_2}^2 E_0 z_0},\nonumber
\end{eqnarray}
and the Ehlers  transformation 
\begin{equation}
E=\frac{E_0}{1+\lambda_E E_0},\quad
\Phi=\frac{\Phi_0}{1+\lambda_E E_0},\quad 
z=z_0+\frac{\lambda_E\Phi_0^2}{1+\lambda_E E_0}. 
\end{equation}

Now let us turn to construction of the infinite hierarchy of potentials.
The advantage of the present formulation consists in the close similarity
to the Kinnersley--Chitre description of the vacuum Einstein equations
(rather than electrovacuum). So we can directly repeat the procedure
in terms of matrix--valued quantities. The first step consists in 
writing down the whole system of equations as a manifestly $G$--covariant
condition of self--duality. To this end one introduces the matrix-valued
twist tensor 
\begin{equation}
{\tilde\nabla}\Psi^A_B =\rho^{-1}\,F^{AC}\,\nabla F_{CB},
\end{equation}
whose existence is implied by the equation (22). 
Taking complex combination
\begin{equation}
H_{AB}=F_{AB}+i\Psi_{AB},
\end{equation}
one can direcly check that it obeys the self--duality condition
\begin{equation}
\nabla H_{AB}=-i\rho^{-1}F_A^C{\tilde\nabla}H_{CB}.
\end{equation}

Infinite hierarchy of potentials ${\stackrel{n}{H}}_{AB}$ 
now can be generated via recursive equations
\begin{equation}
{\stackrel{n+1}{H}}_{AB}=i\left({\stackrel{1n}{N}}_{AB}+ 
H_A^C {\stackrel{n}{H}}_{CB}\right),
\end{equation}
where the qudratic potentials ${\stackrel{mn}{N}}_{AB}$ 
are introduced through the relation
\begin{equation}
\nabla{\stackrel{mn}{N}}_{AB}={\stackrel{m}{H^*}}_{CA}
\nabla{\stackrel{n}{H^C}}_B.
\end{equation}
These relations are valid for all $n,\,m=0, 1, 2,...$ and it
is understood that
\begin{equation}
{\stackrel{0}{H}}_{AB}=i\epsilon_{AB},\quad
{\stackrel{0n}{N}}_{AB} =-i{\stackrel{n}{H}}_{AB}.
\end{equation}

Although formally these relations coincide with those for the vacuum
Einstein equations, an essential complication comes from the fact that 
now the potentials are non--commuting matrices. Hence the problem of 
explicit construction of finite transformations acting on the hierarchy
is much more tedious (details will be given elsewhere).

Alternatively, various
linear deformation problems were suggested to deal with the Einstein
equations, which can be probed in the present case too. We briefly
describe here the formulation of the Belinskii--Zakharov inverse 
scattering transform method  
appropriate to the EMDA system. It can be derived starting with the
null--curvature representation of the equations of motion following
from the symmetric space nature of  
three--dimensional $\sigma$--model target space.  Such representation
was derived in \cite{g} in terms of the $P,\,Q$ variables. Now
we are in a position to give similar formulation directly in terms
of the initial physical quantities entering the Eqs. (22).
 
Rewriting the system (22) as the modified chiral equation for the
symmetric $4\times 4$ matrix $F=F_{AB}$
\begin{equation}
(\rho F_{,\rho} F^{-1})_{,\rho} +  (\rho F_{,z} F^{-1})_{,z}=0,
\end{equation}
one can obtain the corresponding Lax pair with a complex
spectral parameter $\lambda$ in the original Belinskii--Zkharov
form:
\begin{equation}
D_1\Psi=\frac{\rho U-\lambda V}{\rho^2+\lambda^2}\Psi,\quad
D_2\Psi=\frac{\rho V+\lambda U}{\rho^2+\lambda^2}\Psi.
\end{equation}
Here $V=\rho F_{,\rho}F^{-1},\; U=\rho F_{,z}F^{-1}, \; \Psi$
is a matrix "wave function", and
\begin{equation}
D_1=\partial_z-\frac{2\lambda^2}{\rho^2+\lambda^2}\partial_{\lambda},\quad
D_2=\partial_{\rho}+\frac{2\lambda \rho}{\rho^2+\lambda^2}\partial_{\lambda}
\end{equation}
are commuting operators; then the nonliner system (52) can be regarded
as the compatibility condition of the linear system $[D_1, D_2]\Psi=0$.
Similar linear system may be written in terms of the dualized
variables, i.e. the matrix $M$ satisfying Eq. (33).

An infinite algebra of the Geroch--Kinnersley--Chitre group can 
be obtained from here via an expansion of the linear system in power
series in terms of the complex spectral parameter $\lambda$.
Practical use of the $4\times 4$ matrix Lax pair requires further
study. We note another $4\times 4$ linear problem (with different
group structure) discussed recently \cite{lb} 

\vskip20pt
{\bf Acknowledgements} \vskip20pt

The author thanks the organizers of this Conference for kind
hospitality and support. Useful discussions 
with L. Bordag are gratefully acknowledged. At different stages of
this research the author profited much
from conversations with G.A. Alekseev, G. Neugebauer, D. Kramer, 
and J.B. Griffiths. The research was supported 
in part by the Russian Foundation for Fundamental Research 
Grant 93--02--16977, the International Science Foundation and 
Russian Gov. Grant M79300 and the INTAS Grant 93-3262.

\end{document}